\documentclass{PoS}

\usepackage[reqno]{amsmath}
\usepackage{ulem}
\usepackage{amssymb}
\newcommand{\bea}{\begin{eqnarray}}
\newcommand{\eea}{\end{eqnarray}}
\newcommand{\ba}{\begin{array}}
\newcommand{\ea}{\end{array}}

\newcommand{\ovl}{\overline}

\newcommand{\al}{\alpha}
\newcommand{\be}{\beta}
\newcommand{\lam}{\lambda}

\newcommand{\im}{{\rm Im}\,}
\newcommand{\re}{{\rm Re}\,}

\newcommand{\mup}{\mu^{\prime}}

\newcommand{\dmsol}{\mbox{$\Delta m^2_{\odot}$}}
\newcommand{\dma}{ \mbox{$\vert\Delta m^2_{\rm A}\vert$}}

\title{Natural Seesaw Realization at the TeV Scale}

\ShortTitle{Natural seesaw realization at the TeV scale}

\author{\speaker{Emiliano MOLINARO}\\
       Technische Universit\"at M\"unchen\\
       James-Franck-Stra\ss{}e, 85748 Garching, Germany\\
       E-mail: \email{emiliano.molinaro@tum.de}}

\abstract{We summarize the  phenomenological constraints on  seesaw scenarios defined at the TeV
scale and provide a simple extension of the Standard Model which naturally leads to a testable mechanism of neutrino mass generation.}

\FullConference{The European Physical Society Conference on High Energy Physics \\
		 18-24 July, 2013\\
		 Stockholm, Sweden}

\begin{document}
We review the most important phenomenological constraints on type I/inverse seesaw scenarios defined at the electroweak
scale. We also consider a simple extension of the Standard Model (SM) which realizes a ``testable'' seesaw scenario, without imposing
any fine-tuning of the neutrino Yukawa couplings in order to generate light active neutrino masses.
\section{Phenomenological constraints on the TeV scale seesaw parameter space}

We consider a phenomenological  seesaw  \cite{seesaw} extension of the SM with two heavy Majorana fermion singlets $\mathcal{N}_{1,2}$, 
which in principle can  be tested in low energy and collider experiments for masses $M_{1,2}\sim(100-1000)$ GeV.

Following \cite{Ibarra:2010xw}, $\mathcal{N}_{1,2}$ have charged-current (CC) and neutral-current (NC) interactions with the SM leptons and 
interact with the Higgs boson. They are given by
\begin{eqnarray}
 \mathcal{L}_{CC}^\mathcal{N} &=& -\,\frac{g}{2\,\sqrt{2}}\,
\bar{\ell}\,\gamma_{\alpha}\,(RV)_{\ell k}(1 - \gamma_5)\,\mathcal{N}_{k}\,W^{\alpha}\;
+\; {\rm h.c.}\,
\label{NCC},\\
 \mathcal{L}_{NC}^\mathcal{N} &=& -\frac{g}{4 \,c_{w}}\,
\overline{\nu_{\ell L}}\,\gamma_{\alpha}\,(RV)_{\ell k}\,(1 - \gamma_5)\,\mathcal{N}_{k}\,Z^{\alpha}\;
+\; {\rm h.c.}\,,\label{NNC}\\
\mathcal{L}_{H}^\mathcal{N} &=& -\frac{g \,M_{k}}{4\, M_{W}}\,
\overline{\nu_{\ell L}}\,(RV)_{\ell k}\,(1 + \gamma_5)\,\mathcal{N}_{k}\,h\;
+\; {\rm h.c.}\,
\label{NH}
\end{eqnarray}
The couplings $(RV)_{\ell k}$ ($\ell=e,\mu,\tau$ and $k=1,2$) arise from the mixing between heavy and light Majorana neutrinos and, therefore, are suppressed by the seesaw scale. They can be conveniently parametrized as follows~\cite{Ibarra:2010xw}:
 \begin{eqnarray}
\label{mixing-vs-y}
\left|\left(RV\right)_{\ell 1} \right|^{2}&=&
\frac{1}{2}\frac{y^{2} v^{2}}{M_{1}^{2}}\frac{m_{3}}{m_{2}+m_{3}}
    \left|U_{\ell 3}+i\sqrt{m_{2}/m_{3}}U_{\ell 2} \right|^{2}\,,
~~~~{\rm for~normal~hierarchy}\,,\\
\left|\left(RV\right)_{\ell 1} \right|^{2}&=&
\frac{1}{2}\frac{y^{2} v^{2}}{M_{1}^{2}}\frac{m_{2}}{m_{1}+m_{2}}
    \left|U_{\ell 2}+i\sqrt{m_{1}/m_{2}}U_{\ell 1} \right|^{2}\,,
\,~~~~{\rm for~inverted~hierarchy}\,,
\label{mixing-vs-yIH}\\
(RV)_{\ell 2}&=&
\pm i\, (RV)_{\ell 1}\sqrt{\frac{M_1}{M_2}}\,,~\ell=e,\mu,\tau\,,
\label{rel0}
\end{eqnarray}
%
where $U$ denotes the PMNS neutrino mixing matrix and $v\simeq174$ GeV. 
The relative mass splitting of the two heavy Majorana neutrinos 
must be very small, $|M_{1}-M_{2}|/M_{1}\ll 1$, due to the current upper limit on the effective Majorana mass probed in neutrinoless double beta decay experiments \cite{Ibarra:2010xw}. In this case, the flavour structure of the neutrino Yukawa couplings is fixed by the neutrino oscillation parameters \cite{Ibarra:2010xw,Raidal:2004vt} and the two heavy Majorana neutrinos form a pseudo-Dirac fermion. The parameter $y$ in the expressions above represents the largest
eigenvalue of the matrix of the neutrino Yukawa couplings: $y^{2}v^{2}\;=\;2\,M_{1}^{2}\,(\left| (RV)_{e1} \right|^{2}+
\left| (RV)_{\mu1} \right|^{2}+\left| (RV)_{\tau1} \right|^{2})$.

Electroweak precision data (EWPD) provide an upper bound of the size of the Yukawa coupling $y$ for a given seesaw scale, that is $y\leq 0.06\left(M_{1}/100~\text{GeV}\right)$ \cite{Ibarra:2010xw}.
\begin{figure}
\begin{center}
\begin{tabular}{cc}
\includegraphics[width=7.5cm,height=6.5cm]{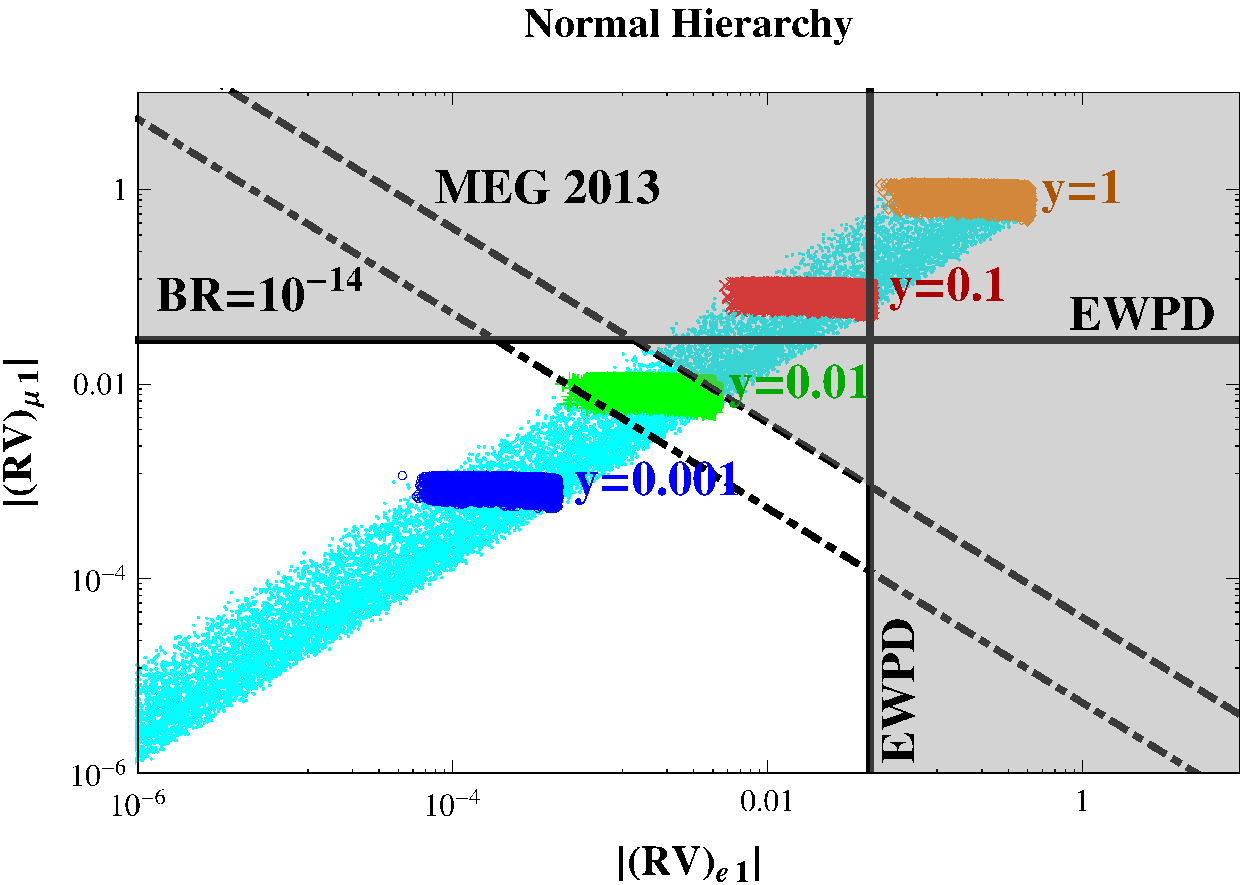} &
\includegraphics[width=7.5cm,height=6.5cm]{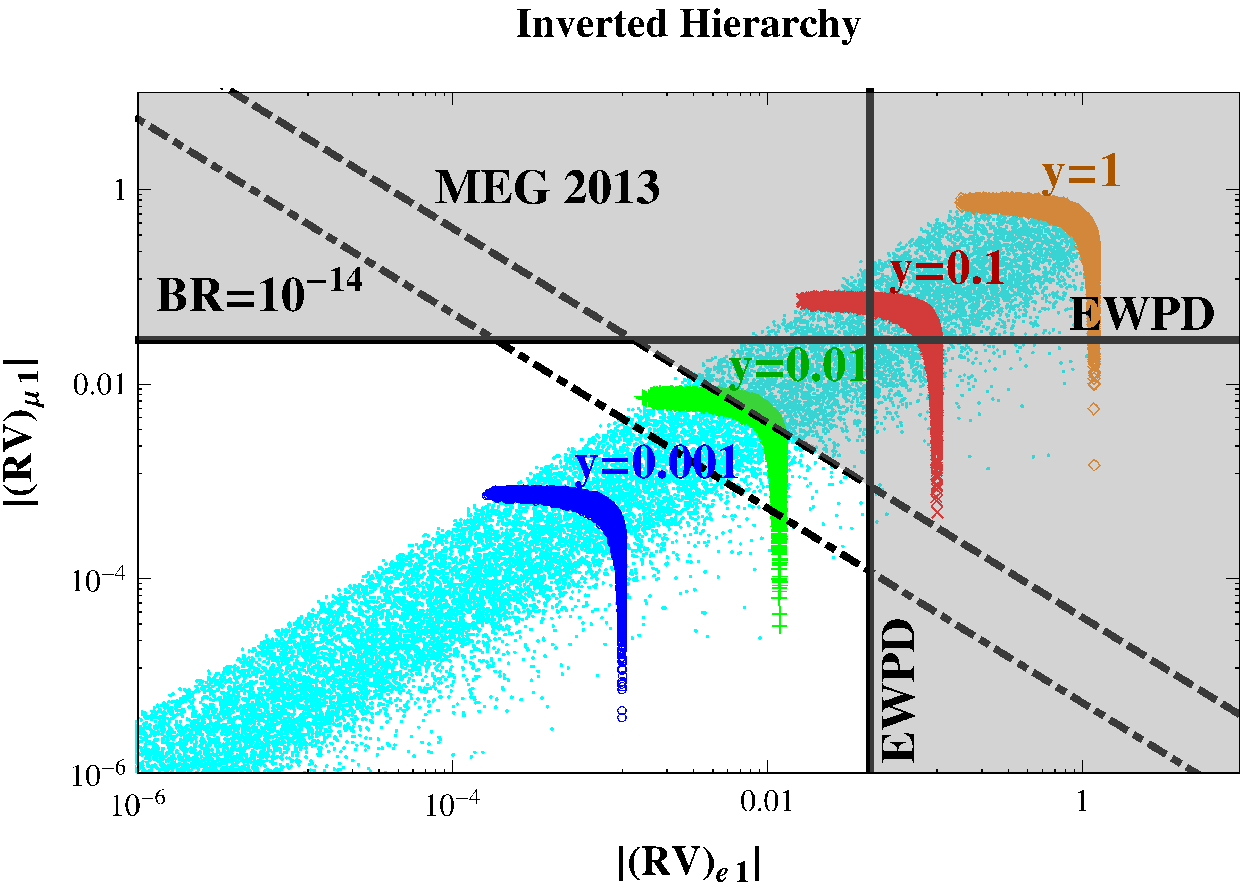}
\end{tabular}
\caption{Correlation between $|(RV)_{e1}|$ and $|(RV)_{\mu1}|$ for 
$M_{1}=100$ GeV in the case of normal (left panel)
and inverted (right panel) light neutrino mass spectrum
(see the text for details). \label{fig1}}
\end{center}
\end{figure}
However, the most stringent constraint on $y$ comes from lepton flavour violating observables, in particular from the present upper limit on $\mu^+\to e^+\, \gamma$ 
branching ratio reported by the MEG experiment \cite{MEG}: ${\rm BR}(\mu\to e\, \gamma)<5.7\times 10^{-13}$ at 90\% confidence level.
Indeed, in this case, taking the best fit values of the neutrino oscillation 
parameters \cite{Tortola:2012te}, we get the bound:  $y\lesssim 0.026$ for $M_1=100$ GeV.

We show in Fig.~\ref{fig1} all the relevant constraints on the effective couplings $(RV)_{\mu 1}$ and $(RV)_{e1}$ which come from the requirement of reproducing neutrino oscillation data, from the EWPD  bounds and the current upper limit on $\mu\to e\,\gamma$. The seesaw mass scale is fixed at benchmark value $M_1=100$ GeV.
From Fig.~\ref{fig1}, it is manifest that the
allowed ranges of the right-handed (RH)
neutrino couplings $|(RV)_{\mu1}|$ and $|(RV)_{e 1}|$, 
in the case of normal (left panel) and inverted (right panel) light neutrino mass spectrum, are confined in a small strip of the overall representative plane. 
This corresponds to the scatter plot of the points 
which are consistent with the $3\sigma$ allowed
ranges of the neutrino oscillation parameters \cite{Tortola:2012te}.
The region of the parameter space 
which is allowed by the EWPD
is marked with solid lines. The region allowed
by the current bound on the $\mu\rightarrow e\,\gamma$ 
decay rate is indicated with a dashed line, while the dot-dashed line shows the exclusion limit for ${\rm BR}(\mu\to e\,\gamma)< 10^{-14}$. 
The scatter points correspond to different values of 
the maximum neutrino Yukawa coupling $y$: 
$y=0.001$ (blue $\circ$), $ii)$
$y=0.01$ (green $+$), $iii)$ $y=0.1$ (red $\times$), $iv)$ $y=1$ (orange $\Diamond$) and 
$v)$  an arbitrary value of the Yukawa coupling $y \leq 1$ (cyan points).

As depicted in Fig.~\ref{fig1}, in the case of a light neutrino mass spectrum with inverted hierarchy a strong suppression of the $\mu\to e\,\gamma$ decay rate is possible for specific values of the measured neutrino oscillation parameters. This is due to cancellations in the $\mu-e$ transition amplitude proportional to $|U_{\mu 2}+i\,U_{\mu 1}|$ which are possible  in the case
of neutrino mass spectrum with inverted hierarchy for $\sin\theta_{13}\gtrsim 0.13$ and CP conserving phases of the neutrino mixing matrix (see \cite{Ibarra:2010xw} for a detailed discussion).

\section{A natural model realization of the TeV scale seesaw scenario}

We discuss a simple model that naturally provides a testable seesaw scenario where the RH neutrino interactions in (\ref{NCC}-\ref{NH}) are ``sizable''
and give rise to observable effects. We extend the scalar sector of the SM with an additional Higgs doublet, $H_{2}$, and a complex 
singlet, $\phi$. The fermion sector, instead, consists of the two RH neutrinos $N_{1,2}$, which allow to implement the seesaw mechanism in order to explain
active neutrino masses and mixing. The resulting model Lagrangian is invariant under a global $U(1)_{X}$ symmetry, where $X$ corresponds to a generalization
of the lepton number. The particle content of the model and the charge assignment of the fields  are reported in Table~\ref{FieldAssign}.
\begin{table}[!t]
\begin{center}
\begin{tabular}{|c||c|c|c|c||c|c|c|c|}
\hline 
\rule[0.15in]{0cm}{0cm}{\tt Field}& $L_{\alpha}$ & $e_{R\alpha}$ & $N_{1}$ & $N_{2}$ & $H_{1}$ & $H_{2}$ & 
$\phi$ \\
\hline
$\rm SU(2)_{\rm L}$ & $\mathbf{2}$ & $\mathbf{1}$ & $\mathbf{1}$ & $\mathbf{1}$&  $	\mathbf{2}$ & $\mathbf{2}$ & 
$\mathbf{1}$ 
\\\hline
$\rm U(1)_{\rm Y}$ & -$1/2$ & -$1$ & $0$ & $0$ & $1/2$ & $1/2$ & 
$0$ 
\\\hline
$\rm U(1)_{\rm X}$ & -$1$ & -$1$ & -$1$  & +$1$& $0$ & $2$ & 
-$2$ 
\\\hline
\end{tabular}
\caption{Charge assignment of the fields.}\label{FieldAssign}
\end{center}
\end{table}%

In this scenario, the presence of $H_{2}$ and $\phi$ is motivated by the requirement of generating light neutrino masses through 
the type I/inverse/linear seesaw mechanism at the TeV scale. The most general scalar potential $\mathcal{V}_{\rm SB}$,  invariant under the SU$(2)_{\rm L}\times$U$(1)_{\rm Y}\times$[U$(1)_{\rm X}$] 
symmetry, is derived in~\cite{Previous}. Given the fields in Table~\ref{FieldAssign}, we get
\bea
\mathcal{V}_{\rm SB}&=&-\mu_{1}^2\,(H_1^{\dag}\,H_1) + \lam_{1}\, (H_1^{\dag}\,H_1)^2 - \mu_{2}^2\,( H_2^{\dag}\,H_2) + \lam_{2}\, (H_2^{\dag}\,H_2)^2 - \mu_{3}^2\, \phi^{*} \phi + \lam_{3}\, (\phi^{*} \phi)^2\nonumber \\  &+& \kappa_{12}\, (H_1^{\dag}\,H_1) \,(H_2^{\dag}\,H_2) +\kappa_{12}^{\prime}\, (H_1^{\dag}\,H_2)\,(H_2^{\dag}\,H_1)+\kappa_{13}\, (H_1^{\dag}\,H_1) \phi^{*} \phi + \kappa_{23}\, (H_2^{\dag}\,H_2) \phi^{*} \phi \nonumber \\
&-&\frac{\mup}{\sqrt{2}}\, \left( (H_1^{\dag}\,H_2) \phi + (H_2^{\dag}\,H_1) \phi^{*}\right)\,.\label{VSB}
\eea
The two SU$(2)_{\rm L}$ doublets $H_{1,2}$ and the singlet $\phi$~\cite{Previous} take a non-zero vacuum expectation value (vev) $v_{1,2}$ and $v_{\phi}$, respectively.
In this case, the global $\rm U(1)_{\rm X}$ is spontaneously broken down to a $Z_{2}$ discrete symmetry.
The scalar  mass spectrum of the model consists of: 1 charged scalar $H^\pm$, 3 CP-even neutral scalars $h^0$, $H^0$ and $h_A$, 2 pseudo-scalars $A^0$ and $J$.
The latter is the Goldstone boson associated with the breaking of the global ${\rm U(1)_{\rm X}}$ symmetry and is usually dubbed Majoron in theories with spontaneously broken lepton charge. Since it is a massless particle, strong constraints apply on its couplings to the SM fermions:  a hierarchical pattern for the vevs of the scalar fields, namely $v_2 \ll v_{1}, v_{\phi}$,  is required in order to satisfy the astrophysical constraints on the Majoron phenomenology. As discussed in detail in~\cite{Previous}, a suppressed value of $v_{2}\propto \mup$ is naturally 
realized from the minimization of the potential (\ref{VSB}), due to the residual symmetries of the model.

In the limit of negligible $v_2$, the longitudinal gauge boson components are  $W_{L}^\pm\sim H_1^\pm$ and $Z_L \sim \sqrt{2}\,\im{(H_1^0)}$, while the scalar mass eigenstates are to a good approximation: $H^\pm \sim H_2^\pm$, $h_A \sim \sqrt{2}\;\re{(H_2^0})$, $A_0 \sim \sqrt{2}\;\im{(H_2^0)}$ and $J \sim \sqrt{2}\; \im{(\phi)}$. Moreover, the two neutral scalars $h^{0}$ and $H^{0}$ arise from the mixing of  $\sqrt{2}\;\re{(H_{1}^{0})}$ and $\sqrt{2}\;\re{(\phi)}$.
Typically, we have $v_2 \lesssim 10$ MeV \cite{Previous}, $v_1\simeq 246.2$ GeV, while $v_\phi$ is free.
Recalling that only $H_1$ has Yukawa couplings to SM fermions (cf. Table~\ref{FieldAssign}), $h_A$, $A^0$ and $H^\pm$ couple to the SM sector only through gauge interactions and via the scalar quartic couplings, while $h^0$ and $H^0$ can have \textit{a priori} sizable Yukawa couplings to SM fermions (see \cite{Previous} for a discussion of the collider constraints on the scalar sector of the model).

\subsection*{Neutrino mass generation}
We introduce for convenience a Dirac fermion field, $N_{D}\equiv P_{R}\,N_{1}+P_{L} \,N_{2}^{C}$, where $P_{L,\,R}$ are the usual chiral projectors and $N_{2}^{C}\equiv C \overline{N_{2}}^{T}$ is
the conjugate of the $N_{2}$ RH neutrino field.
The most general interaction Lagrangian of $N_{D}$ invariant under the global $U(1)_{X}$ symmetry is
\bea
\mathcal{L}&\supset& -\,m_{N}\,\ovl{N_{D}}\,N_{D} - \left( Y_{\nu 1}^{\beta}\, \ovl{N_{D}} \,\widetilde{H}_1^{*}\,L_\beta\,+ Y_{\nu 2}^{\gamma} \,\ovl{N_{D}}^{C}\, \widetilde{H}_2^{*}
\,L_\gamma+\frac{\delta_N}{\sqrt{2}}\,\phi\,\ovl{N_{D}}\,N_{D}^{C}  \,+\,{\rm h.c.}\right)\label{neutrL}
\eea
where  $N_{D}^{C}\equiv C\overline{N}_{D}^{T}$ and $\widetilde{H}_{k}\equiv -i\sigma_{2}H_{k}^{*}$ ($k=1,2$). 
The parameter $\delta_N$ is made real through a phase transformation.
The Yukawa interactions $\propto Y_{\nu 1}$ ($Y_{\nu 2}$) couple $N_1$ ($N_{2}$) to the SM leptons. Therefore, 
after the Higgs doublets acquire a nonzero vev, the SM lepton number (i.e. the generalized $X$ charge) is explicitly violated by $Y_{\nu 2}$ mediated interactions.
Furthermore, while the Dirac type mass $m_{N}$ conserves the lepton number, 
the term proportional to $\delta_N$ provides, after $\phi$ takes a nonzero vev, a Majorana mass term for both $N_1$ and $N_2$.
In the case $m_N\gg \delta_N\,v_\phi$ we have a low energy realization of the type I/inverse seesaw scenario \cite{Previous}. 

In the chiral basis $\left(\mathbf{\nu_{ L}}\;, (N_{1}^{C})_{\bf L},\, (N_{2}^{C})_{\bf L} \right)$, the $5\times 5$ symmetric neutrino mass matrix  reads:
\bea
\mathcal{M}_{\nu}=
\left(
\begin{array}{ccc}
\mathbf{0_{3\times3}} & \mathbf{y_{1}}\,v_1 & \mathbf{y_{2}}\,v_2\\
\mathbf{y_{1}}^{\rm T}\,v_1 & \delta_N\,v_\phi & m_{N}\\
\mathbf{y_{2}}^{\rm T}\,v_2 & m_{N} & \delta_N\,v_\phi
\end{array}
\right)\,.\label{massnu}
\eea
In the previous expression $\mathbf{0_{3\times3}}$ denotes the null matrix of dimension 3 and we introduce the shorthand notation: $\mathbf{y_{k}}\equiv \left( Y_{\nu k}^{e}\;,Y_{\nu k}^{\mu}\;,Y_{\nu k}^{\tau}\right)^{\rm T}/\sqrt{2}$.
The neutrino sector, therefore, consists of one massless neutrino, two massive light Majorana neutrinos and two heavy Majorana neutrinos $\mathcal{N}_{1,2}$. The latter are quasidegenerate, with masses $M_{1,2}=m_N\mp\ \delta_N\,v_{\phi}$, and form a pseudo-Dirac pair for $m_N\gg \delta_N\,v_\phi$  \cite{Previous}. They have, therefore, naturally ``sizable'' CC and
NC interactions with the SM leptons, eqs.~(\ref{NCC}) and (\ref{NNC}), where in this case the mixing matrix elements $(RV)_{\ell 1}$ are proportional to the lepton number conserving Yukawa couplings
$Y_{\nu k}^{\ell}$.
The resulting effective light neutrino mass matrix is
\bea
(M_\nu)^{\al \beta}\simeq -\frac{v_1\,v_2}{m_N}\,\left(\mathbf{y_{1}}^{\alpha }\,\mathbf{y_{2}}^{\beta}+\mathbf{y_{2}}^{\alpha }\,\mathbf{y_{1}}^{\beta}\right)+v_\phi\,\delta_N\,\frac{v_1^2}{m_N^2}\left(\mathbf{y_{1}}^\al \mathbf{y_{1}}^\be+\mathbf{y_{2}}^\al\, \mathbf{y_{2}}^\be \,\frac{v_2^2}{v_1^2}\right)\,.\label{massnu4}
\eea
The first term in (\ref{massnu4}) acts as a linear seesaw contribution and its suppression originates from the small vev $v_2$.
The second term is typical of inverse seesaw scenarios,  
where the small ratio $v_\phi\,\delta_N / m_N$ suppresses the neutrino mass scale. 
Notice that, with only  two RH neutrinos  the linear seesaw contribution alone (i.e. neglecting $v_{\phi}$ in (\ref{massnu4}))
allows to fit  all current neutrino oscillation data, while
if $v_{2}=0$ and $v_{\phi}\neq 0$ the inverse seesaw scenario can only account for one massive light neutrino.
Therefore, the complex scalar field $\phi$, with vev $v_{\phi}\neq 0$,  is not mandatory in order to obtain two massive light neutrinos through the (linear) seesaw mechanism.
On the other hand, $v_{\phi}\neq 0$ is a necessary condition to set a hierarchy between the Higgs doublet vevs, $v_{2}\ll v_{1}$, without fine-tuning of the parameters \cite{Previous}.
Taking $\mu_N \ll 1$, the neutrino masses have a simple expression:
\bea
m_\nu^\pm &\simeq & \frac{1}{m_N}\left(\sqrt{ y_1^2\,y_2^2-\mu_N\,(y_1^2+y_2^2)\,{\rm Re }(y_{12})}	\pm \sqrt{ |y_{12}|^2-\mu_N\,(y_1^2+y_2^2)\,{\rm Re }(y_{12})}\right) \nonumber \\
&\simeq& \frac{1}{m_N}\left(\, y_1\,y_2\pm \vert y_{12}\,\vert\right)\times\left(1 \mp \frac{\mu_N}{2}\frac{(y_1^2+y_2^2)\,{\rm Re}(y_{12})}{y_1\,y_2\,\vert y_{12} \vert}\,\right)\,,
\eea
with  $y_i\equiv\sqrt{\mathbf{y_{i}}^\dagger\cdot\mathbf{y_{i}}}\, v_i $, $y_{12}\equiv\mathbf{y_{1}}^\dagger\cdot\mathbf{y_{2}} \, v_1\,v_2$, $\eta_{12}\equiv\mathbf{y_{1}}\times \mathbf{y_{2}}\, v_1\,v_2$ and $\mu_N= (\delta_N\,v_\phi)/m_N$.
Notice that if the neutrino Yukawa vectors  $\mathbf{y_{1}}$ and
 $\mathbf{y_{2}}$ are proportional, $m_{\nu}^{-}$ is exactly zero. 
For a normal hierarchical spectrum, $m_\nu^+= \sqrt{\dma}$ and $m_\nu^-=\sqrt{\dmsol}$, while in the case of  inverted hierarchy 
we have $m_\nu^+=\sqrt{\dma}$ and $m_\nu^-=\sqrt{\dma-\dmsol}$, $\dma$ and $\dmsol$ being the atmospheric and solar neutrino mass square differences, respectively.
It is easy to show 
 that at leading order in $\mu_N$, the neutrino mass parameters satisfy the  relation
\cite{Previous}: $\left| \mathbf{y_{1} \times y_{2}}  \right|\,v_{1}\,v_{2}/m_{N} \cong (\dmsol\,\dma)^{1/4}$.
Hence, barring accidental cancellations, the size of
the neutrino Yukawa couplings is typically
\begin{equation}
|\mathbf{y_{1}}| |\mathbf{y_{2}}|\approx 10^{-4}\,(m_{N}/1~\text{TeV})\,(1~\text{KeV}/v_{2})\,.\label{yuk12}
\end{equation}
Finally we point out that with the addition of a scalar field odd under $U(1)_{X}$ it is possible to realize a viable dark matter candidate in the model,
which results naturally stable due to the presence of the remnant $Z_{2}$ symmetry. A variation of leptogenesis in this case is possible 
at the TeV scale \cite{Previous}.\\

\noindent This work was supported by the ERC Advanced Grant project ``FLAVOUR''(267104).

\end{document}